\documentclass[jcp,amsmath,amssymb,preprint]{revtex4-1}

\usepackage{color}
\usepackage{graphicx}
\usepackage{dcolumn}
\usepackage{bm}

\usepackage{siunitx}
\usepackage{multirow}
\usepackage{cancel}
\usepackage{ulem}

\begin{document}

\preprint{}

\title{Helium bubbles in liquid lithium: a potential issue for ITER}

\author{Edgar Álvarez-Galera, Jordi Martí, Ferran Mazzanti}%
\altaffiliation{Department of Physics, Polytechnic University of Catalonia-Barcelona Tech, B4-B5 Northern Campus UPC, 08034 Barcelona, Catalonia, Spain}
\email{jordi.marti@upc.edu (corresponding author)}
\author{Lluís Batet}
\email{lluis.batet@upc.edu}
\altaffiliation{Department of Physics, Polytechnic University of Catalonia-Barcelona Tech, Diagonal 647, 08028 Barcelona, Catalonia, Spain}

\date{\today}

\begin{abstract}
{Future fusion nuclear reactors will produce sustainable energy form the fusion of deuterium and tritium.  In order to do so, the reactors will need to produce their own tritium through the neutron bombardment of lithium.  Such reaction will produce tritium and helium inside the breeding blanket of the reactor.  Helium can trigger nucleation mechanisms due to its very low solubility inside liquid metals.  Consequently,  the knowledge and understanding of the microscopic processes of helium nucleation is crucial to improve the efficiency,  sustainability and safety of the fusion energy production.  The formation of helium bubbles inside the liquid metal used as breeding material may be a serious issue that has yet to be fully understood.  We provide further insight on the behavior of lithium and helium mixtures at experimentally corresponding operating conditions (800~K and pressures between 1 and 100 bar) using a suitable microscopic model able to describe the helium and lithium atomic interactions, in excellent agreement with available experimental data. The simulations predict the formation of helium bubbles with radii around $\SI{10}{\angstrom}$ at ambient pressure and with surface tension values between 0.6-1.0 N/m,  with a 
dependency of the concentration of helium.  We also report cohesive energies of helium as well as a quantitative estimation of the Hildebrand and Kumar cohesion parameters. } 
\end{abstract}

\maketitle

\section{\label{intro} Introduction} 
As promising providers of clean energy, fusion nuclear reactors were conceptualised and some prototypes were proposed several decades ago.  Actually the ITER project\cite{iter-web},  designed back in 1998\cite{aymar2002iter} but under continuous development is expected to create a first plasma in 2025.  Beyond this,  the following step in fusion, i.e. the first demonstration power plant (project DEMO) is already under study and design\cite{federici2019overview}.  The design and development of such facilities, still in progress, represents one of the main challenges of the modern scientific community.  Sustainable reactors must generate more energy than the one required to activate all nuclear fusion reactions that have to take place within it, i.e., there must be a net energy production. Nevertheless, the self-sufficiency of energy is not the only requirement since tritium must be produced within the reactor itself. 

The nuclear fusion reaction with the highest cross section, i.e., the easiest to produce in a reactor, involves deuterium (D) and tritium (T) as reactants, leading to the production of helium plus a free neutron that escapes at the same time that 17.6 MeV of energy are released\cite{kordavc2017helium}:

\begin{equation} 
        {\rm D} + {\rm T} \rightarrow \;^{4}{\rm He} + {\rm n}+ 17.6\; {\rm MeV}.
    \label{eq0}
\end{equation}
While water is a source of deuterium, tritium must be ``breeded" within the so-called breeding blankets (BB) of the reactors.  Its production, therefore, must be ensured by the reactor itself (self-sufficiency requirement).  Tritium is produced from neutron capture by any of the two natural isotopes of lithium $^6$Li and $^7$Li, with relative abundances of $7.59 \%$ and $92.41 \%$, respectively\cite{rubel2019fusion}:

\begin{equation} 
        {\rm n} + \;^{6}{\rm Li} \rightarrow  {\rm T} + \;^{4}{\rm He} +  4.78\;{\rm MeV}
    \label{eq1}
    \end{equation}
    \begin{equation} 
        {\rm n} + \;^{7}{\rm Li} \rightarrow  {\rm T} + \;^{4}{\rm He} 
        + {\rm n} -  2.47\;{\rm MeV}.
    \label{eq2}
\end{equation}
Some BB designs include a liquid metal (LM) cooled by helium or water. The lead-lithium eutectic (LLE) is one of the main proposals. The addition of lead to the lithium required for the breeding is due to safety reasons, since pure lithium is extremely reactive.  Helium, which is produced in same amounts that tritium (see Eq.\ref{eq2}), as a noble gas is hardly soluble in lithium and other alkali metals. During the tritium breeding, nucleation of helium atoms may occur, affecting thermo-physical properties of the BB and having a great impact in their designs. The solubility of helium in lithium was experimentally measured long time ago, estimating a Henry's constant around $10^{-14}$-$10^{-13}\SI{}{\pascal^{-1}}$ for temperatures well above the melting point of lithium\cite{slotnick1965solubility}. Engineers need to characterise the onset of nucleation of helium in LLE in order to design the proper PbLi circuits.  

Computer simulations allow to study the described phenomenon. In this framework, we simulate the nucleation of helium as nano-bubbles at several thermodynamic conditions using the molecular dynamics (MD) method.  MD is used to model and simulate a wide variety of systems, ranging between liquid hydrogen to large parts of cells, by means fo the generation of Newtonian trajectories in phase space,  which alows us to computed a enormous collection of properties from thermodynamics to classical and quantum dynamical properties (free energies,  transport coefficients or permittivities, to cite only a few\cite{padro1994molecular, calero20151h}). It provides also a direct route to match key experimental data such as structure factors or atomic and molecular spectra\cite{brunger1987crystallographic,padro2004response} in order to validate the models employed in the simulations. Since the key point of the possible nucleation in BB is assumed to arise from helium-lithium and helium-lead interactions, we consider first binary lithium-helium systems previous to the study of the behavior of the noble gas in LLE. In the present work we investigate the lithium-helium repulsion in full details.  We have assumed the choice of the Belaschenko's Embedded Atom Model (EAM)\cite{belashchenko2012embedded} together with the Toennies-Tang-Sheng (TTS) model\cite{sheng2020conformal,sheng2021development}, which has lead to segregation of both species. Specially, when the concentration of helium is considerably low ($x_{\rm He}=\frac{N_{\rm He}}{N}\leq 0.04$) the repulsion exerted by the presence of larger amounts of lithium atoms in their surroundings makes helium atoms gather in pseudo-spherical clusters with radii within the nanoscale.

In a previous work\cite{marti2022nucleation}, lithium-helium mixtures were simulated using the Neutral Pseudo-Atom (NPA) potential for lithium-lithium interactions, proposed by Canales et al.\cite{canales1994computer} for pure lithium, and specific Lennard-Jones (LJ) pair potentials for helium-lithium and helium-helium interactions,  parameterized by the authors.  With that model, the onset of nucleation was clearly established in a qualitative fashion.  Despite of the accuracy of many structural properties in the canonical (NVT) ensemble that set of potentials overestimated forces, leading to poor estimations of properties such as surface tension.  Furthermore,  an increasing stability of helium clusters with pressure was found, against the overall behavior described by Henry's law for gases dissolved in liquids.  Other authors like Fraile et al.\cite{fraile2020volume} have faced the problem of the nucleation of helium in lead-lithium alloys using an EAM for both lead and lithium potential terms of the Hamiltonian. However, where these authors used LJ potentials for those terms that involve helium atoms, in this work we rely on TTS models, since the latter have softer and more realistic repulsive regions and are expected to be more capable to reproduce properties like solubility\cite{sheng2021development}. We realized that LJ potentials overestimate the repulsion between helium and lithium atoms for very short distances, although they are very similar to TTS models for intermediate and long distance ranges. Usually potential models are validated by comparing with experimental data quantities related to pure lithium,  such as densities, diffusion coefficients, Hildebrand's or Kumar's parameters and radial distribution functions.  In the present work, other quantities related to helium-in-lithium systems are also investigated: size of the helium clusters, their cohesion energy and the interfacial tension between the two phases. This paper is organized as follows: in Sec \ref{methods} we present the microscopic model and the computational details for the simulations.  In Sec \ref{res} we verified the models, study the size of helium clusters of atoms (radii and fraction of atoms that join to the largest cluster),  report the binding potential energy\cite{marti2022nucleation} as an estimation of the cohesion of these clusters and explain how we compute forces, liquid and gas pressures and the surface/interfacial tension. Finally, in Sec \ref{concl}, we summarize all the key points and remarks of this work, discussing about future investigations that can be continued from the actual state of the research. 
    
\section{\label{methods} Models and methods}
    \subsection*{\label{mmod} Pairwise interaction potentials}

We describe a classical system of particles whose dynamics is driven by the Hamiltonian, 
\begin{eqnarray}
        \label{eq_hamilt}
        H & = & \frac{1}{2}\sum\limits_{i=1}^{N_{\rm Li}} m_{\rm Li}v_{{\rm Li}, i}^2
        +\frac{1}{2}\sum\limits_{i=1}^{N_{\rm He}}m_{\rm He}v_{{\rm He}, i}^2\\
        &+& \sum\limits_{i=1}^{N_{\rm Li}} \Phi_{\rm Li} (\rho_i) + \sum\limits_{i<j}^{N_{\rm Li}}V_{{\rm Li}-{\rm Li}}(|{\bf r}_{{\rm Li},i}-
        {\bf r}_{{\rm Li},j}|) \nonumber \\
        &+&\sum\limits_{i<j}^{N_{\rm He}}V_{{\rm He}-{\rm He}}(|{\bf r}_{{\rm He},i}-
        {\bf r}_{{\rm He},j}|) \nonumber  \\
        &+& \sum\limits_{i=1}^{N_{\rm Li}}
        \sum\limits_{j=1}^{N_{\rm He}}
        V_{{\rm Li}-{\rm He}}(|{\bf r}_{{\rm Li},i}-{\bf r}_{{\rm He},j}|)\;.
        \nonumber
\end{eqnarray}
We fix $N_{\rm Li}$ and $N_{\rm He}$ as the number of lithium and helium atoms, respectively,  being $N=N_{\rm Li}+N_{\rm He}$ the total number of atoms.  Atoms are described as massive point-like particles,  but we will also refer to the atomic Van der Waals radius $\sigma$, that corresponds to the atom-atom distance where the corresponding interaction potential is zero.  The Hamiltonian is composed by: (1) the kinetic energy of Li and He species; (2) the embedding energy of lithium $\Phi_{\rm Li}$, which is a function of the atomic electron density seen by the \textit{i}-th Li atom $\rho_i = \sum_{j \neq i} \psi (|{\bf r}_{{\rm Li},i}-{\bf r}_{{\rm Li},j}|) $\cite{daw1983semiempirical} and (3) pair interaction potentials: $V_{\rm Li-Li},  V_{\rm He-He}$ and $V_{\rm Li-He}$ as described in the "Supporting Information" (SI).  As usual,  particle coordinates and velocities of each species $\{ {\bf r}_{{\rm Li},i}, {\bf v}_{{\rm Li},i} \}$ and $\{ {\bf r}_{{\rm He},j}, {\bf v}_{{\rm He},j} \}$ determine instantaneous configurations.  Although we consider a classical description of the system, interaction models are determined by their quantum nature, including the interaction of lithium atoms with the electron cloud by means of the electronic density and the Pauli exclusion at very short distances due to overlapping of electron orbitals. The potentials used in this work also include a highly non-monotonic behavior at distances near to the Van der Waals radii and an attractive long-range tail.

Liquid lithium is modelled using the Daw-Baskes\cite{daw1983semiempirical} formalism, relying on the EAM model proposed by Belashchenko in Ref.~\cite{belashchenko2012embedded}. The explicit form of functions the pairwise interactions $\varphi(r_{ij}) \equiv \varphi_{ij} \equiv V_{{\rm Li}-{\rm Li}}(|{\bf r}_{{\rm Li},i}-{\bf r}_{{\rm Li},j}|)$, $\psi_{ij} \equiv \psi(r_{ij})$ and functional $\Phi_{i} \equiv \Phi(\rho_{i})$ (embedding function of the electrons) is reported in Ref.\cite{belashchenko2012embedded}, but we also provide a detailed description in SI. The two kind of interactions that involve helium atoms are modeled with TTS potentials\cite{sheng2020conformal,sheng2021development},  which were constructed from first-principles quantum calculations. The TTS model is known to provide a good description of rare gas hydrides and alkali-helium systems, in excellent agreement with previous {\it ab-initio} calculations of Partridge et al.\cite{partridge2001potential}, who found that theoretical values of Li–He total scattering cross sections and the rare-gas atom–He transport properties agree well (to within about 1\%) with the corresponding experimental data.  More specifically, the helium-lithium interaction potential provided by the TTS model is able to reproduce with high precision the quantum-mechanical fits provided by Partridge et al. The full list of coefficients for each TTS potential model considered in this work has been summarized in SI, together with the appropriate formulas to obtain intermolecular forces.  There we can observe that a given set of parameters is taken for He-He and Li-He.  In particular, a number $N_{\rm max}=8$ of terms in the expansion ensures the convergence of the series for both He-He and Li-He (5 is enough for the former,  but 8 is more appropriated to reproduce the softer repulsive regions of the Li-He interactions), while larger numbers could introduce errors due to the overuse of the recurrence relation that determine the dispersion coefficients \cite{sheng2021development}.  

\begin{figure}[htbp]
      \begin{center}
                \centering
                \includegraphics[width=0.8\columnwidth]{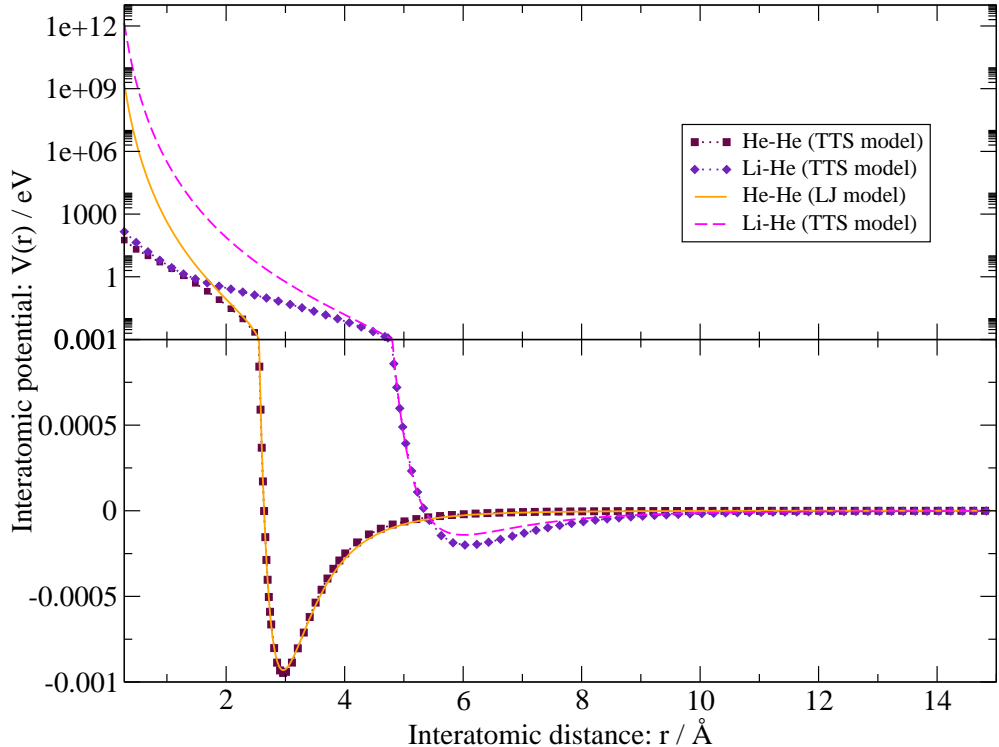}
                \caption{Comparison of TTS and LJ potentials for interactions involving helium atoms (He-He and Li-He). The top part is represented in logarithmic scale and it shows the repulsive part of the potential (short-distance range). The bottom part shows the long range (van der Waals) region.}
                \label{fig:TTS_vs_LJ}
        \end{center}        
      \label{fig:TTS_HeHe}
\end{figure}

For the sake of completeness of the description of the interaction models used in this work, we add a comparison 
of TTS and the widely employed Lennard-Jones models for helium interactions, in Figure \ref{fig:TTS_vs_LJ}.  As a 
noble gas, the interactions of helium are very well represented by LJ potentials at intermediate and long distance 
ranges.  In particular, the potentials constructed from the TTS model for helium-helium and helium-lithium 
interactions are very similar to the de Boer-Michels LJ potential function\cite{de1938contribution,aziz1987new} for distances above $\SI{2}{\angstrom}$ and to the Dehmer-Wharton potential function \cite{dehmer1972absolute} for distances above 
$\SI{5}{\angstrom}$. We can observe that beyond the highly repulsive region below $\SI{2}{\angstrom}$ both LJ and TTS potentials have very similar location ($2^{1/6}\sigma^{\rm LJ}$) and depth of their minima ($-\varepsilon^{\rm LJ}$). This similarity has been tested through statistical analysis of all the inter-atomic distances. We found that there are no recorded distances below $\SI{2}{\angstrom}$ for helium-helium and below $\SI{5}{\angstrom}$ for lithium-helium, at moderate pressures of the order 1 bar and temperatures up to 20\% over the fusion point, found at 415 K for the systems reported in this work. However, in the case of the large pressures and temperatures that may be of interest in fusion technology, the similarity between TTS and LJ is no longer true,  arising possible inaccuracies in the determination of the super-saturation threshold, where a simulation involving 40 helium and 1000 lithium atoms usinf LJ models would require a pressure of $10^6\SI{}{\bar}$\cite{marti2022nucleation}.  While other authors considered the LJ model for lithium-helium interactions\cite{fraile2020volume} we rely on the TTS potential because of its more accurate quantum nature and its reliability at low distances.  A more 
detailed discussion on the choice of the TTS model over LJ potentials will be given in a forthcoming paper, in which 
Henry's constant of helium-in-lithium will be calculated using the Frenkel's et al.  cavity scheme\cite{li2017computational,li2018computational,wand2018addressing,espinosa2018calculation}.
        
\subsection*{\label{cd} Computational details}
        
We assumed the Hamiltonian described in Eq.~\ref{eq_hamilt} within the MD scheme. The integration of the equations of motion is performed with the \textit{LAMMPS} package\cite{LAMMPS}.  The simulations are performed in the following way: we first consider fixed temperature, $T$ and number of lithium atoms, $N_{\rm Li}$.  Lithium atoms are placed within a cubic simulation box of length $L$, which must be compatible with experimental values of the density of lithium, $\rho_{\rm exp, Li} (T)$\cite{zinkle1998summary,davison1968compilation}. Standard periodic boundary conditions (PBC) have been applied.  Lithium atoms can be located at random within the box or located at the nodes of a lattice with spacing $a = L / N^{1/3} = \left( \rho_{\rm exp, Li}(T)\right)^{-1/3}$.  In order to mimic the technological conditions of the BB in the fusion reactor,  the concentration of helium atoms must be much lower than that of lithium, so that we consider atomic fractions of helium $x_{\rm He}=\frac{N_{\rm He}}{{\rm N}}\leq 0.075$. 

\begin{figure}[htbp]
            \begin{center}
                \includegraphics[width=0.8\columnwidth]{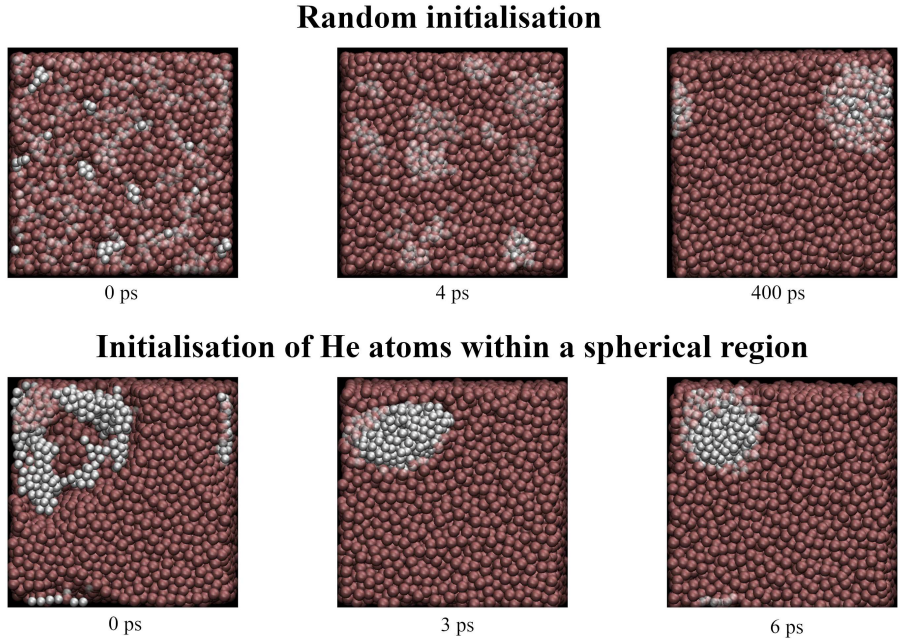}
                \caption{Comparison of initialisation and cluster formation ($N_{\rm He}$) when all atoms are initially located at random (top) and when helium atoms are generated at random within a spherical region (bottom). In both cases the system evolves towards similar equilibrium states, where all the atoms gather in a single spherical-like cluster.}
                \label{fig:evol}
            \end{center}
\end{figure}

We have considered two ways of initialisation before equilibration: (1) helium atoms initially placed randomly inside the simulation box and (2) helium atoms located in a spherical region inside the box. In both cases the system evolved towards equilibrium configurations in which a single helium bubble embedded inside the lithium solvent appeared (see Figure \ref{fig:evol}), with no smaller clusters observed as a general fact.  As it can be observed from this figure, the second method ensures a much faster way to equilibrium, since He atoms are generated close within a same region, and around 6 ps instead of the 400 ps required if helium atoms are distributed at random.  Before performing equilibration runs, the conjugate gradient (CG) algorithm\cite{hestenes1952methods} is applied to avoid possible overlaps or unstable configurations.  For the sake of a quick generation of the initial configurations, during CG and the first equilibration steps we use LJ potentials with the parameters shown in table \ref{tab_LJ}, since the use of the EAM and TTS models is significantly more expensive,  requiring much longer computational times. 

\begin{table}
            \begin{tabular}{c|c c }
                \toprule
                Interaction & $\epsilon$ (K) &  $\sigma$ (\AA) \\ \hline \hline
                \textbf{Li-Li}\cite{canales1994computer,marti2022nucleation} &  833.6 & 2.800 \\\hline
                \textbf{He-He}\cite{de1938contribution,aziz1987new} & 10.22 & 2.556 \\ \hline
                \textbf{Li-He}\cite{dehmer1972absolute,fraile2020volume} & 1.636 & 5.3565 \\
                \hline \hline
            \end{tabular}
            \caption{Values of the parameters used to model LJ pairwise interactions.}
            \label{tab_LJ}
\end{table}

These LJ potentials correspond to a preliminary model of the mixture: while He-He and Li-He are extracted directly from Refs.\cite{aziz1987new,dehmer1972absolute}, respectively, Li-Li arises from a fit to the NPA potential of Canales et al.\cite{canales1994computer}, also employed in a preliminary work\cite{marti2022nucleation}.  Once a reasonable distribution of 
atoms is generated, interactions are turned into the ones of Belashchenko\cite{belashchenko2012embedded} and 
Toennies-Tang-Sheng\cite{sheng2020conformal,sheng2021development}.  Additional equilibration steps are performed 
before starting the production runs in order to collect statistically meaningful properties.  In the production runs
the equations of motion are integrated with by coupling the system to a Nosé-Hoover thermostat/barostat with 
damping parameters of $\SI{0.04}{ps}$ when NVT/NpT configurations are generated.  The integration is performed by the 
velocity-Verlet\cite{swope1982computer} algorithm, with time steps fixed at $\Delta t = \SI{0.002}{ps}$ in all cases.

Once formed, the position of the helium bubble is determined, every 1000 time steps, by calculating center of mass (CoM) of helium atoms,  including PBC,  as it is explained in SI.

\section{\label{res} Results and discussion}

We start reporting the results used to validate the models described above (Section \ref{methods}). In all the cases, the concentration of helium has been set to low values below 7~\%.  We have considered two characteristic system sizes: simulation boxes of $L \sim \SI{30}{\angstrom}$ containing 1000 lithium atoms plus 0 and plus 40 helium atoms and $L \sim \SI{65}{\angstrom}$ containing 10648 lithium atoms and between 0 and 800 helium atoms.  Simulations of large systems were carried out when statistical fluctuations observed in the small ones were too large,  such as radii of the helium bubbles or for surface tension. Furthermore, we made sure that all properties are properly converged. 

\subsection{\label{dd} Densities and diffusion coefficients of lithium}

The density of lithium atoms, computed as $\rho_{\rm Li} = N_{\rm Li} / V$, varies during NpT simulations due to volume fluctuations. The average values of $\rho_{\rm Li}$ at several temperatures are reported in Table \ref{tab_lithium_properties}.  All simulations were totally converged after 1 ns, producing the same average values (within statistical errors) of the lithium density and diffusion coefficient for all temperatures considered.  Our results revealed to be in good agreement with the experimental density of lithium found in Refs.\cite{zinkle1998summary,davison1968compilation}. 

\begin{table}
        \caption{Densities ($\rho_{\rm Li}$),  temperatures ($T$) and diffusion coefficients of lithium ($D_{\rm Li}$) for setups made of 1000 and 10648 lithium atoms in the NpT ensemble.  All MD simulations span a total time length of 1 ns.}
        \centering
        \begin{tabular}{c|c|ccc}
        \toprule
        $N_{\rm Li}$ & $T$ (K) & $\rho_{\rm Li}$ (kg m$^{-3}$)& $D_{\rm Li}$ (\AA$^2$/ps) \\ \hline
        \multirow{5}{*}{1000} & 470 & $500 \pm 30$ & $0.7 \pm 0.4$ \\
        ~ & 563 & $490 \pm 20$ & $1.29 \pm 0.12$ \\
        ~ & 657 & $480 \pm 20$ & $1.7 \pm 0.3$ \\
        ~ & 750 & $470 \pm 20$ & $2.4 \pm 0.3$ \\
        ~ & 843 & $460 \pm 20$ & $3.0 \pm 0.3$ \\ \hline
        \multirow{5}{*}{10648} & 470 & $500 \pm 30$ & $0.7 \pm 0.4$ \\
        ~ & 563 & $490 \pm 30$ & $1.30 \pm 0.13$ \\
        ~ & 657 & $480 \pm 30$ & $1.86 \pm 0.18$ \\
        ~ & 750 & $470 \pm 30$ & $2.45 \pm 0.19$ \\
        ~ & 843 & $460 \pm 30$ & $3.1 \pm 0.3$ \\ \hline\hline
        \multirow{2}{*}{Experiment} & 470 & 513.3 (Ref. \cite{zinkle1998summary})/ 515.0 (Ref. \cite{davison1968compilation}) & 0.64 (Ref. \cite{canales1993molecular}) \\
        & 843 & 480.683 (Ref. \cite{zinkle1998summary}) / 477.7 Ref. \cite{davison1968compilation}) & 2.58 \cite{canales1993molecular})\\\hline \hline
        \multicolumn{1}{c}{~} 
        \label{tab_lithium_properties}
        \end{tabular} 
\end{table}

As observed from the results shown in Table \ref{tab_lithium_properties}, the system sizes considered in ths work are already large enough to produce converged values of both the density of lithium above the melting point,  found at 415 K.  As expected,  densities gradually decrease as temperature increases, given the larger thermal energy of the system, which produces a slight increase of the volume.  

\begin{figure}
    \centering
    \includegraphics[width=0.8\columnwidth]{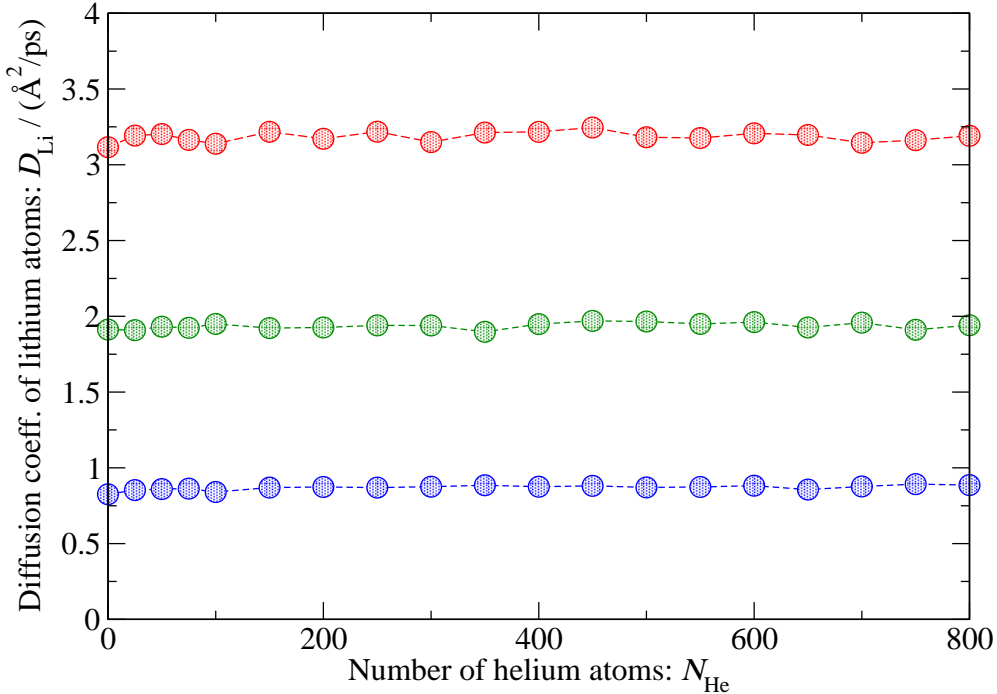}
    \caption{Diffusion coefficients of lithium for simulations as a function of the number of helium atoms for NpT simulations. 470 K (blue circles), 657 K (gree circles) and 843 K (red circles).}
    \label{fig:diffusion}
\end{figure}

For the sake of validation of the dynamics of lithium, we also provide the diffusion coefficient $D$ obtained as a fit to the slope of the $t \rightarrow \infty$ mean-square displacement curve.  For a species $\alpha$, $D_\alpha$ is given by the Einstein's expression valid for Brownian motion:

 \begin{equation}
 D_\alpha = \frac{1}{6}\;\lim_{t\rightarrow \infty}\;\frac{{\rm d}}{{\rm d}t}
                \langle|{\bf r}_i(t)- {\bf r}_i(0)|^2\rangle_{i = 1, ..., N_\alpha},
            \label{eq_einstein}
 \end{equation}
where ${\bf r_i(t)}$ and ${\bf r_i}(0)$ stands for the coordinate of each atom of species $\alpha$ at instants $t\rightarrow\infty$ and 0.  The diffusion coefficient of lithium atoms is reported in Table \ref{tab_lithium_properties}. Their values at temperatures of 470 and 843 K are in overall good agreement with the available experimental ones reported in Ref.\cite{canales1994computer}. 
Not surprisingly,  lithium diffuses faster as the temperature increases. The results reported in Figure \ref{fig:diffusion} show that lithium dynamics is essentially independent of the presence of helium bubbles when the pressure is fixed, in a wide range of helium concentrations between 0\% and 7.5\%. 

\subsection{\label{size} Size of helium clusters}

Assuming spherical symmetry, the size of a cluster of helium atoms can be characterised by its radius or, alternatively, by the number of helium atoms contained in the cluster. In most simulations we observe that He atoms tend to group after thermalization.  Only in very large systems with a low concentration of helium atoms one can observe the formation of small formed by tens of helium atoms.  We denote $N_{\rm cl}$ as the number of helium clusters that emerge, each one containing $N_{\rm He}^k$ ($k = 1, \dots, N_{\rm cl}$) atoms. The principal component, which is the cluster containing the largest amount of helium atoms within the box, is identified with the label $p$ and has a size of $N_{\rm He}^p = \max\{N_{\rm He}^k\}_{k = 1, \dots, N_{\rm cl}} \leq N_{\rm He}$.  Clusters and their size are identified from atomic coordinates and are analysed by means of a geometric-based algorithm. 

The radius of the He clusters have been determined as the square root of mean square distance between He atoms and the their CoM, namely the radius of gyration $R_{\rm MSD}$ of the cluster: 
\begin{equation} \label{eq_rad_MSD}
    R_{\rm MSD} \equiv \sqrt{\langle r^2 \rangle},    
\end{equation}
being $\displaystyle \langle r^2 \rangle = \langle {\bf r}'_{\rm He} \cdot {\bf r}'_{\rm He}\rangle = \frac{1}{N_{\rm He}^p} \sum_{i'=1}^{N_{\rm He}^p}  {\bf r}'_{i'} \cdot {\bf r}'_{i'}$ the average over all He atoms in a single frame and ${\bf r}'_{\rm He} = {\bf r}_{\rm He} - {\bf r}_{\rm CoM}$.  Similarly to Ref.~\cite{cui2015molecular}, inner and outer radius of the helium lithium interface can be determined from the particle density profile, $\rho (r)$, but are not reported here for the sake of clarity.

\begin{table}
\caption{Radius of gyration and surface tension $\gamma_{\rm S}$ of helium clusters (computed using several approaches,  such as Young-Laplace, Thompson and Bakker-Buff,  see Section\ref{forces}),  for setups made of 10648 lithium and 400 helium atoms. All MD simulations were of total length 1 ns and pressure was below 100 bar.}
        \centering
            \begin{tabular}{c|c|ccccc}
            \toprule
            \multicolumn{2}{c|}{T (K)} & 470 & 653 & 657 & 750 & 843 \\ \hline \hline
            ~ & Method &\\ \hline
            \multirow{2}{*}{$R_{\rm MSD}$ (\AA)} & \ref{eq_rad_MSD} & $9.768 \pm 0.012$ & $10.274 \pm 0.015$ & $10.85 \pm 0.03$ & $11.46 \pm 0.03$ & $12.13 \pm 0.04$ \\
            ~ & \ref{eq_rad} & $9.14 \pm 0.03$ & $9.63 \pm 0.02$ & $10.19 \pm 0.03$ & $10.77 \pm 0.02$ & $11.37 \pm 0.04$ \\ \hline
             \multirow{3}{*}{$\gamma_{\rm S}$ (N/m)} & Bakker-Buff & $0.832 \pm 0.003$ & $0.780 \pm 0.004$ & $0.723 \pm 0.006$ & $0.670 \pm 0.007$ & $0.617 \pm 0.007$ \\
            ~ & Thompson & $0.946 \pm 0.003$ & $0.883 \pm 0.003$ & $0.816 \pm 0.005$ & $0.754 \pm 0.007$ & $0.695 \pm 0.006$ \\
            ~ & Young-Laplace & $1.007 \pm 0.004$ & $0.939 \pm 0.003$ & $0.865 \pm 0.005$ & $0.798 \pm 0.006$ & $0.735 \pm 0.005$ \\ \hline \hline
        \end{tabular}
        \label{tab_bubble_properties}
\end{table}        

\begin{figure}[htbp]
\begin{center}
\includegraphics[width=0.8\columnwidth]{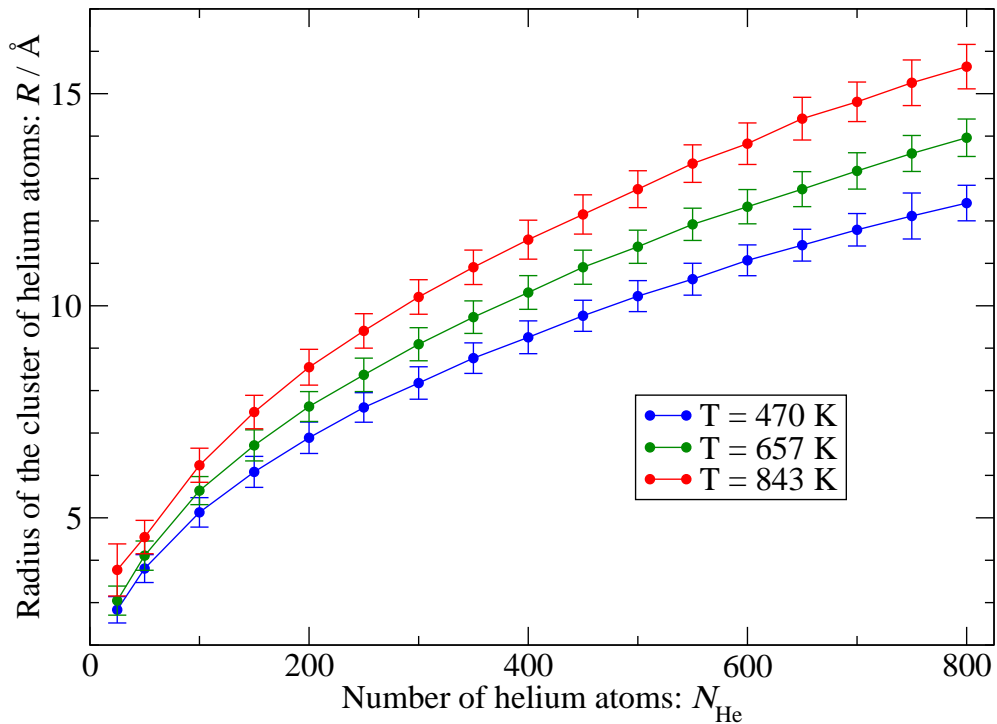}
\caption{\label{fig_RvsNHe} Radii of helium bubbles formed by 10 to 800 helium atoms, with $N_{\rm Li} = 10648$. The pressure was fixed at 1 bar. 470 K (blue circles), 657 K (gree circles) and 843 K (red circles). The dferent criteria described in the text have been averaged, since they provide very similar values (see Table \ref{tab_bubble_properties}.}
\end{center}
\end{figure} 

The results are reported in Table\ref{tab_bubble_properties} for 400 He atoms and as a function of $N_{\rm He}$ in
Figure\ref{fig_RvsNHe}. As a general fact,  we observed that the radii of the He clusters depends on both the temperature and the pressure, although the latter dependency has not studied with detail.  As a function of the temperature,  we find a clear trend of increasing radii with rising temperatures,  producing values between 9 and 12 \AA$\;$ at 470 and 843 K, respectively.

\subsection{\label{gdr} Radial distribution functions}

\begin{figure}[htbp]
\begin{center}
\includegraphics[width=0.8\columnwidth]{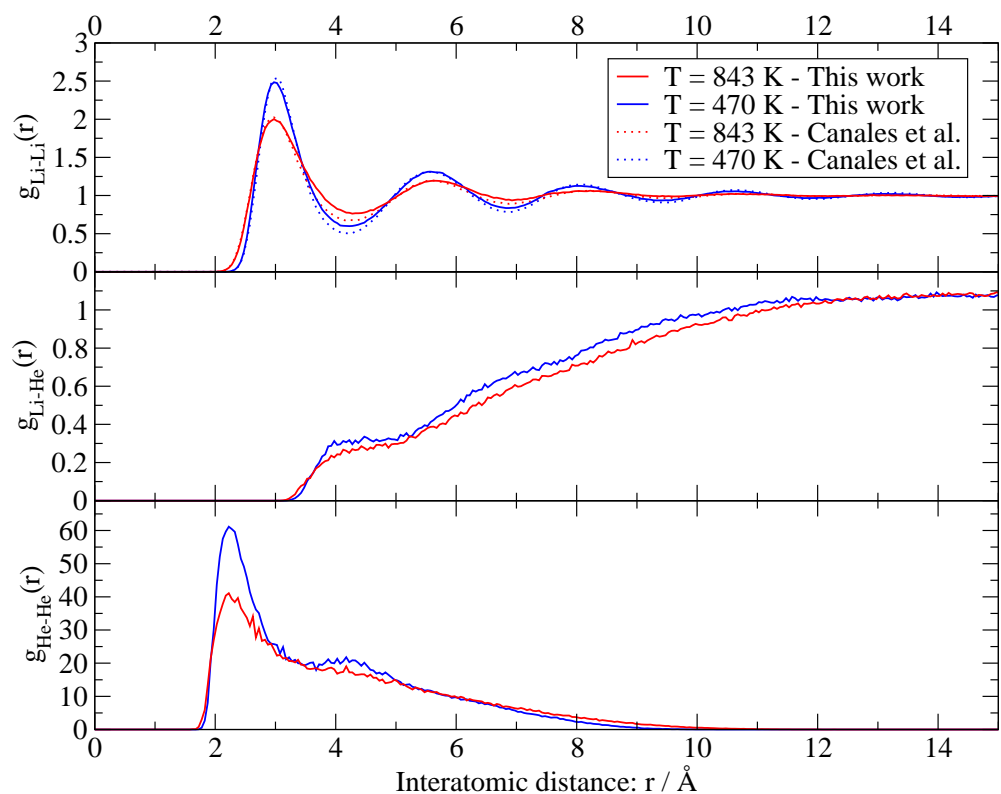}
\caption{\label{fig:gdrs} Li-Li, Li-He and He-He pair distribution functions at temperatures of 470K and 843 K. Dashed lines correspond to experimental RDFs reported in \cite{canales1994computer}, while straight lines correspond to this work, i.e. simulations with the EAM potential for Li-Li interactions and the TTS model for He-He and Li-He.  The simulations have been performed including 1000 Li atoms + 40 He atoms and constant volume.}
 \label{fig_gdr}
\end{center}
\end{figure}

The arrangement of atoms and structural properties are characterized by means of the radial or pair distribution function (RDF), given by:

\begin{equation}
              g_{\alpha - \beta}(r) = \frac{1}{N_\alpha N_\beta}
              \sum_{i=1}^{N_\alpha} \sum_{j=1}^{N_\beta}
              \left\langle \delta( |{\bf r}|_{ij} - r)
              \right\rangle\;,
            \label{eq_gdr}
\end{equation}
being $\langle \cdots\rangle$ a thermal average that denotes the probability that two atoms $i$, $j$, of species $\alpha$ and $\beta$, respectively, are separated a distance $r$ apart, where $\alpha$ and $\beta(r)$ may be lithium or helium. Figure\ref{fig_gdr} shows the characteristic shapes of both $g_{\rm Li-Li}(r)$ and $g_{\rm He-He}(r)$ at temperatures of 470 and 843 K in the NVT ensemble.  In the case of lithium-lithium pairs, both RDF are compared with the experimental ones at same temperatures, reported by Canales et al.\cite{canales1994computer}.  

The Li-Li RDF, $g_{\rm Li-Li}(r)$, reports that lithium behaves as a typical liquid and it shows strong similarity to the pure lithium RDF regardless of the presence of helium atoms (reminding the low helium concentration conditions) i.e.  since $N_{\rm Li} >> N_{\rm He}$, the presence of helium has no great impact on $g_{\rm Li-Li}(r)$ in comparison to pure lithium.  Both $g_{\rm Li-Li}(r)$ at temperatures of 470 and 843 K are in agreement with the ones presented at the same respective temperatures in Ref.~\cite{canales1994computer} and obtained from experimental data.

The tail of a spherical condensate of radius $R$ usually decays to zero as $g_{\rm He-He}(r \sim R) \simeq A \exp{(-2r/R)}$ (quantum bond states).  Consequently,  assuming such approach, the distance where $g_{\rm He-He}(r) \rightarrow 0$ gives a rough estimation of the size of the cluster ($\sim \SI{9}{\angstrom}$ for $T = \SI{470}{K}$ and $\sim \SI{10.5}{\angstrom}$ for $T = \SI{843}{K}$), in overall agreement with the estimations of the radius of the helium bubbles from geometric considerations (see Table\ref{tab_bubble_properties} in Section\ref{size}).
            
Finally,  from the helium-lithium RDF we simply corroborate the strong repulsive interaction between them at short distances 
and some hint of a structure around 4\AA$\;$ and beyond.

\subsection{\label{bind} Cohesive forces: binding potential energy of helium atoms and cohesion parameters of lithium}

In the studies of solubility of gases inside liquid metals, it is very relevant the determination of appropriate solubility parameters.
In the case of helium, the correct assessment of the cohesive forces between helium and the eutectic alloy atoms is crucial\cite{sedano2022solubility} to understand its behaviour and predict new properties.  Usually, the cohesion 
solvent-solvent forces are quantified by a cohesion parameter given by the ratio between specific cohesive energy and the specific volume.  In a similar fashion to previous research~\cite{marti2022nucleation},  it was considered the so called 
binding (or cohesive) energy of helium defined as

\begin{equation} \label{eq_binding}
            U_{\rm binding} \equiv \frac{U_{\rm He-Li} - U_{\rm Li}}{N_{\rm He}},
\end{equation}
where $U_{\rm He-Li} $ and $U_{\rm Li}$ are the internal energies of the mixture and pure lithium, respectively and $N_{\rm He}$ is the total number of helium atoms.  $U_{\rm Li}$ is calculated from an independent simulation of pure lithium ($N_{\rm He} = 0$ and same $N_{\rm Li}$ as in the mixture). The comparative values of the binding energies of helium can be used to justify the stability of a cluster of helium atoms. 

\begin{figure}
        \centering
        \includegraphics[width=.8\columnwidth]{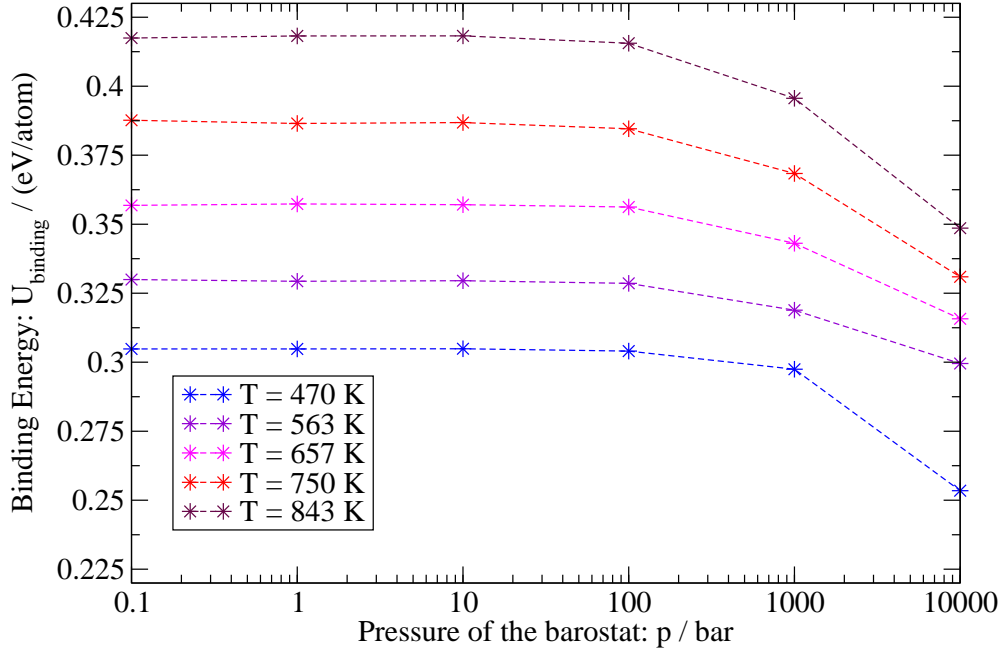}
        \caption{Binding energies as a function of temperature and pressure}
        \label{fig:BPE_pres}
    \end{figure}

Results for the binding energy as a function of the pressure are displayed in Figure~\ref{fig:BPE_pres} for five selected temperatures.  The binding energies, reported in eV/atom show a clear dependence on the temperature, with a tendency to rise with increasing temperatures. On the other hand,  $U_{\rm binding}$ is nearly independent of the pressure below 100 bar, while at high pressures the binding energies decrease moderately.  We should notice that $ U_{\rm binding}$ is related to the chemical potential\cite{pal1994change}. The difference $U_{\rm He-Li} - U_{\rm Li}$ is linked to the Gibbs free energy change, 
$\Delta G$, when inserting $N_{\rm He}$ helium atoms.  A change of the sign while increasing pressure may indicate a reverse tendency to aggregation: negative for large pressures (when nucleation occurrs i.e., helium atoms bound) and positive for small ones (when dissociation takes place, i.e., helium atoms not bound). In a previous work~\cite{marti2022nucleation}, it was observed using NPA\cite{canales1994computer} and LJ models that clusters of helium atoms were more stable at large pressures, a result that seems to be against Henry's law.

    \begin{figure}
        \centering
        \includegraphics[width=.8\columnwidth]{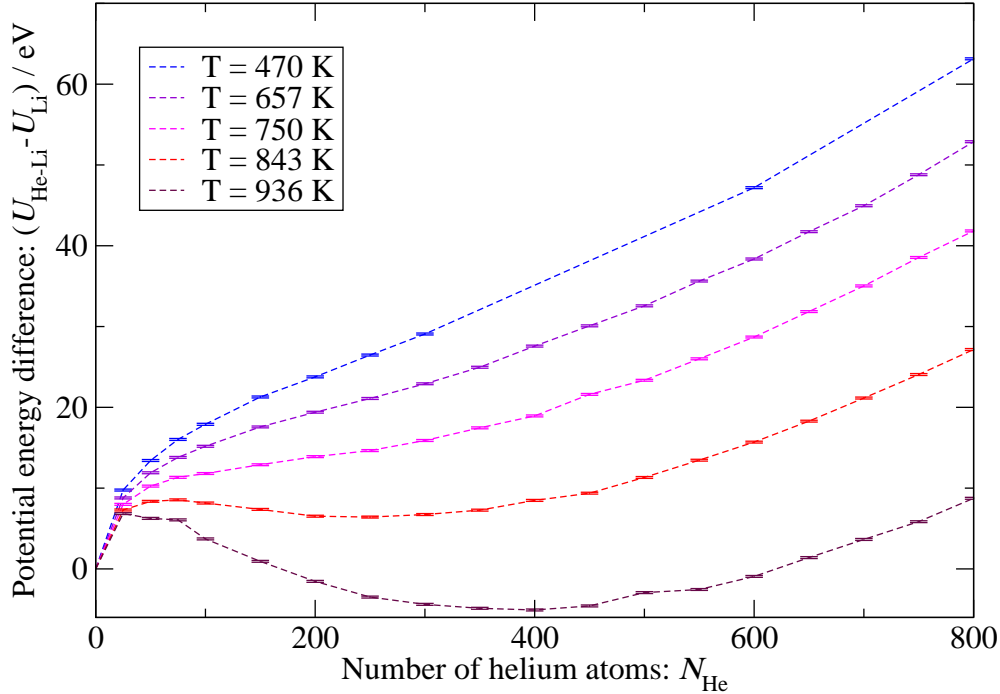}
        \caption{Difference between the average potential energy  $U_{\rm He-Li}$ of systems of $N_{\rm He}$ helium atoms and systems without helium atoms  $U_{\rm Li}$ (fixing 10648 lithium atoms) as a function of $N_{\rm He}$. }
        \label{fig:BPE_vs_N}
    \end{figure}
In Figure~\ref{fig:BPE_vs_N} we represent the dependence of the difference of potential energies $U_{\rm He-Li}^{\rm pot} - U_{\rm Li}^{\rm pot}$ with helium's concentration at five selected temperatures.  We can observe the outbreak of a minimum at a certain temperature between 750 K and  843 K.  For temperatures below that threshold the trend monotonically increasing, whereas at 843 and 936 K we can observe such a minimum for around 400 helium atoms (i.e.  3.7\% concentration) that might indicate the number of helium atoms giving rise to the energetically most stable bubbles, given the  relationship of $U_{\rm He-Li} - U_{\rm Li}$ with the Gibbs free energy sketched above.  

To complete the study on binding energies, we report in SI $U_{\rm binding}$ as a function of the number of helium atoms when simulating the NVT and NpT ensembles.  We find that, in the canonical ensemble, $U_{\rm He-Li}^{\rm pot} - U_{\rm Li}^{\rm pot}$ presents a relative minimum at $x_{\rm He} \sim 3\%$, while in the isothermal-isobaric representation, $U_{\rm He-Li} - U_{\rm Li}$ is decreasing with $x_{\rm He}$.  This indicates a tendency to stabilization, found as helium concentration increases in the NVT simulations, reaching a minimum at a concentration of 2.8\% corresponding to a negative value of $U_{\rm binding}$.  Conversely,  when the pressure is fixed to values around 1 bar no minima is found,  with binding of helium atoms becoming more stable as concentration increases.  Hence the existence of a concentration of helium giving best stability of the bubbles is only hinted around 3\%, but only at large temperatures above 843 K.
  
As a matter of fact, the cohesion forces (or cohesion energies) can be parameterized according to  different approximations reported in the literature.  The two most outstanding are that of Hildebrand\cite{hildebrand1954simple} and of Kumar\cite{kumar1972immiscibility}.  We have verified the similarity of the Hildebrand's cohesion parameter value for pure lithium ($\delta_{\rm H}$), which is defined as

\begin{equation}  \label{Hilde}
\delta_{\rm H} = \sqrt{\frac{\Delta H - k_{\rm B}T}{V}} = \sqrt{\frac{-U_{\rm Li}^{\rm pot}}{V}},
\end{equation}
where $\Delta H$ is the heat (enthalpy) of vaporization of lithium and $k_B$ is the Boltzmann's constant.  The equality 
$-U_{\rm Li}^{\rm pot} = \Delta H - k_{\rm B}T$ stands for the cohesion energy associated to the attractive interaction 
of liquid lithium,  since kinetic energies vanish when the system undertakes a phase change (vaporization) at constant temperature and it is valid for temperatures well below the lithium boiling point (experimentally found around 1608 K\cite{davison1968compilation}), as the ones considered in the present work. 

The intramolecular forces acting within condensed phases lead to negative potential energy values $U_{\rm Li}$, so that $-U_{\rm Li}>0$.  Consequently,  from formula\ref{Hilde} we have estimated the Hildebrand's cohesion parameter for pure lithium as $\delta_{\rm H} = (98 \pm 2) \text{ MPa}^{1/2}$,  which it is consistent with the values of $\SI{98.7}{MPa^{1/2}}$ and $\SI{112}{MPa^{1/2}}$ reported in Refs.\cite{sedano2022solubility,barton2017crc}, respectively.  

Finally,  Kumar's cohesion parameter $\delta_{\rm K}$ is computed similarly to Hildebrand's one,  but considering the enthalpy of melting instead of the one of vaporisation\cite{sedano2022solubility}.  Again,  from the fundamental relationship $\displaystyle dH = T dS + V dP = dU + d(pV)$, we compute the change of enthalpy as $\Delta H|_{\rm NVT} = \Delta U + V \Delta p$ in the canonical ensemble and $\Delta H|_{\rm NpT} = \Delta U + p \Delta V$ decreasing gradually the temperature from a state at which lithium is known to be liquid until it solidifies.  For both NVT and NpT simulations with 1000 lithium atoms we obtain a value  $\delta_{\rm K} \simeq \SI{17.2}{MPa^{1/2}}$, in good agreement with the value reported by Sedano\cite{sedano2007helium} of 16.2 $\text{MPa}^{1/2}$. These two parameters are widely regarded as benchmarks for the reliability of the modeling of liquid lithium-helium mixtures\cite{sedano2022solubility}.  
         
\subsection{\label{forces} Interatomic forces: pressure of liquid and gas phases and surface tension of the helium bubbles}

In order to estimate the stability of the helium bubbles, we consider helium to be in a gas phase within the liquid metal. Then 
we can use several approaches valid for the computation of the surface tension of the He bubbles.  The Young-Laplace (YL) capillarity equation\cite{thompson1984molecular},

\begin{equation}
            \gamma^{YL} = \frac{\Delta p R_{\rm bubble}} {2},
            \label{eq_laplace}
\end{equation}
provides a first approach from the difference of pressure between the liquid metal (lithium) and the gas (helium 
bubble) $\Delta p = p_{\rm LM} - p_{\rm bubble}$ and the radius of the bubble ($R_{\rm bubble}$).  However, because 
of its poor atomistic description the validity of this approach to compute the surface tension in the nano-scale has been discussed\cite{caro2015capillarity,park2001molecular}.  In this scheme,  both the pressure of the gas and the density are functions of the position and cannot be defined as global quantities from an equation of state\cite{caro2013structure} i.e.  far from the constant pressure's assumption of the YL equation.  Conversely, the method by Thompson et al.\cite{thompson1984molecular} provides an atomistic path to the surface tension. This method, often referred as a MD calculation of surface tension,  computes this magnitude from the first derivative of the normal component of the pressure tensor
        
\begin{equation} \label{eq_pN}
p_N (r) = k_B T \rho (r) - \frac{1}{4 \pi r^3} \sum _k |{{\bf r} \cdot {\bf r}_{ij}}|\frac{1}{r_{ij}} \frac{du(r_{ij})}{dr_{ij}}.
\end{equation}
In Eq. \ref{eq_pN}, $\rho(r)$ is the density profile with origin at the CoM, obtained as the number of helium atoms in spherical shells of thickness $\Delta r$,
\begin{equation}
\rho (r) = \frac{N(r)}{4 \pi (r^2 + \frac{\Delta r}{12})\Delta r}.
\end{equation}
This method assumes spherical clusters of atoms,  which lead to a pressure tensor of the form $\Bar{\bar{p}}({\bf r})=p_N(r)\hat{e}_r^T\hat{e}_r + p_T(r)[\hat{e}_\theta^T \hat{e}_\theta+\hat{e}_\phi^T \hat{e}_\phi]$. Assuming the general mechanical equilibrium condition, $\nabla \vec{p} = 0 $ and using mechanical arguments for forces and torques on a hypothetical strip cutting the surface of the drop\cite{thompson1984molecular}, we obtain the surface tension integrating $\displaystyle r^3 \frac{dp_N(r)}{dr}$ over the variable $r$, which is the distance of a given point inside the simulation box respect to the CoM of the bubble.

\begin{figure}
            \centering
            \includegraphics[width=0.8\columnwidth]{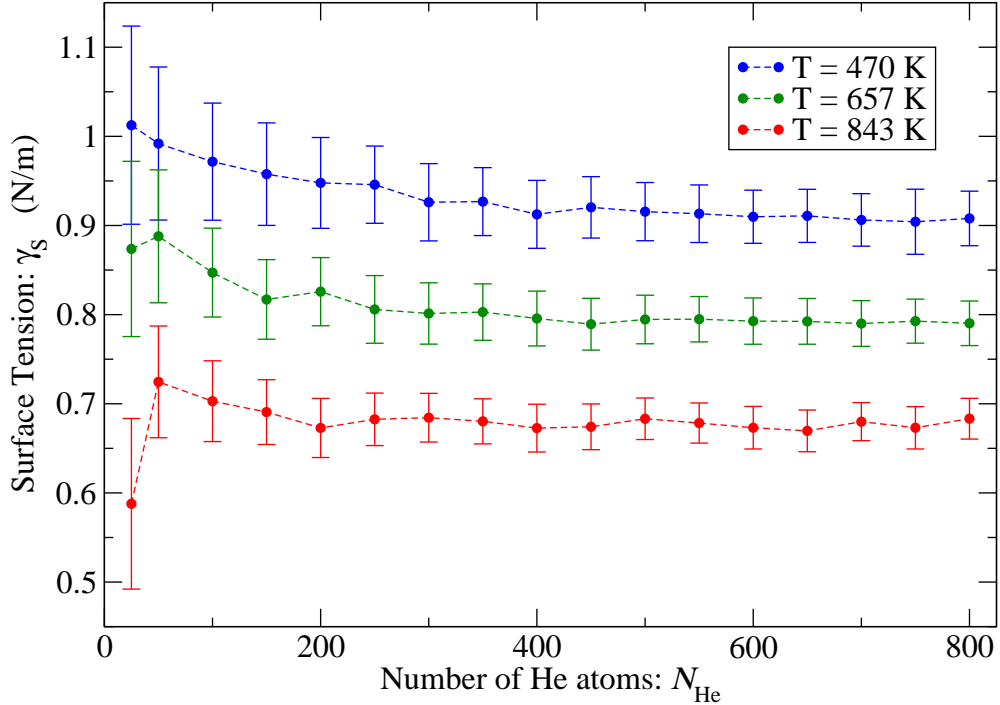}
            \caption{Estimation of the surface tension of helium bubbles using Thompson's method.  470 K (blue circles), 657 K (gree circles) and 843 K (red circles).}
            \label{fig:ST}
\end{figure}

Because of the large fluctuations of pressure-related quantities, pressure components and the profile $p_N(r)$ are averaged from several equilibrium configurations (snapshots) of the simulation, in which clusters are stable (i.e. they have a fixed number of helium atoms). Finally, the surface tension is calculated as

\begin{equation}
\gamma_{\rm S} = \left[ -\frac{1}{8} \Delta p^2 \int_0^\infty r^3 \frac{dp_N(r)}{dr} dr \right]^{1/3}.
            \label{eq_ST_Thompson}
\end{equation}
Although Eq.~\ref{eq_ST_Thompson} was originally developed to study the surface tension of liquid drops\cite{thompson1984molecular}, it has also been tested the relative validity on micro-bubbles\cite{park2001molecular} 
and nano-bubble surface tension of LJ fluid-solid systems\cite{rezaei2011molecular}.  Using this method we have obtained the surface tension of the He bubbles as a function of their size. The results using Eq.\ref{eq_ST_Thompson}
are displayed in Fig.\ref{fig:ST} while their averaged numerical values are reported in Table~\ref{tab_bubble_properties} for the three approaches considered and for $N_{\rm He} = 400$.  We have observed that, because of the wide fluctuations of the pressure, the uncertainty on the value of the surface tension is considerably large.  As a general trend, we have found quite a marked dependence on the temperature with larger surface tension for lower temperatures and an overall slight tendency of increasing $\gamma_{\rm S}$ as the radius of the bubble decreases. However, due to the uncertainty in the difference of pressures $\Delta p$, the dispersion of values becomes larger for small He bubbles.  On the other hand, the Young-Laplace approach overestimates the average value of the surface tension, and their relative uncertainties are as large as their values. Interestingly, the average value of the YL surface tension tends to converge to a constant value with increasing $N_{\rm He}$, at the same time that the dependence on temperature is clearer than in the MD case.
    
Given the large fluctuation in the computation of $\Delta p$, we can also use the equation obtained by Bakker and Buff\cite{thompson1984molecular} that, assuming spherical symmetry and mechanical equilibrium, can be written as 

\begin{equation}
    \gamma_{\rm S} = -\frac{1}{2 R_{\rm S}^2} \int_0^\infty r^3 \frac{dp_N(r)}{dr} dr,
    \label{eq_ST_Bakker}
\end{equation}
where $R_{\rm S}$, that is the radius of the surface of the helium cluster,  estimated using the mean squared distance criterion. Notice that Eqs.~\ref{eq_ST_Thompson} and \ref{eq_ST_Bakker} are equivalent when the Young-Laplace equation is also valid. However, we see that both Thompson's et al.  and Young-Laplace estimations of the surface tension are both larger than the one predicted using \ref{eq_ST_Bakker}, being Young-Laplace values mostly overestimated.  Since obtaining $R_{\rm S}$ is a more accurate task providing better results than calculating directly $\Delta p$, we consider the values obtained using the Bakker-Buff equations as the best estimation of $\gamma_{\rm S}$.  For the reference case of $N_{\rm He} = 400$, when using Eq.~\ref{eq_ST_Bakker} we get 0.597~N/m at 843~K, 0.723~N/m at 657~K and 0.823~N/m at 470 K.  These values are systematically smaller than the average values reported in Fig.~\ref{fig:ST} using Thompson's approach (Eq. \ref{eq_ST_Thompson}).  For the sake of comparison with measured data,  we can consider the experimental values of the surface tension of lithium\cite{davison1968compilation,zinkle1998summary} which are around 0.3-0.4 N/m, i.e. of the same order of magnitude as the reported values of the He bubbles.  Furthermore,  the experimental surface tension of helium ranges between 0.35 N/m at ultra-low temperatures around 1K,  down to 0.1 N/m at the boiling point around 4K\cite{wohlfarth2008surface}.  No data has been found at higher temperatures. 

The three approaches described above lead to linear dependence of the surface tension on the temperature, $ \gamma (T) = 
A + B T$, in the range that has been studied. The parameters of the linear fits are reported in SI (Table\ref{linfits}).  As a final remark,  we can estimate a certain surface energy of a cluster using Eq.~\ref{eq_ST_Bakker},  as the value given by 
$\displaystyle 4 \pi \gamma R^2 = -2\pi \int_0^\infty r^3 \frac{dp_N(r)}{dr} dr$.  Such energy is displayed in Figure \ref{fig:INTER_ENER} of SI and indicate similar energies for small bubbles and a tendency to become larger as the size of the bubbles increase.

 \section{Concluding remarks}
\label{concl}

In this work we have analysed the thermodynamics, structure and dynamics of lithium-helium mixtures with a very low He concentration as a first step towards the simulation of the typical environmental conditions in the BB of a fusion power plant.  We performed classical simulations of the lithium-helium mixture using suitable potential models (TTS and EAM) and molecular dynamics methods, which allowed us to obtain densities, binding energies, and to characterize its structure through the RDF and time-dependent quantities such as the diffusion coefficients of lithium. The validation of the potential models and simulation procedures has been achieved comparing our predictions with available experimental data. Further, we are reporting specific properties of the Li-He mixture in fusion processes such as the Hildebrand and Kumar parameters, and microscopic characteristics of the He bubbles such as their radii and surface tension.

We have seen that, given the low helium solubility in the range of pressures considered in the present work, helium bubbles are formed in a wide range of temperatures and pressures, so that independently of the initial homogeneous disposition of atoms in the system, our simulations showed the formation of helium for the same environmental conditions. Furthermore, we observe that helium atoms are miscible in the lithium bath at low pressures. The size of the helium bubbles has been found to be in the range of 10~\AA, while the surface tension of the bubbles has been estimated to be between 0.6-1.0 N/m.  Future studies would likely involve the calculation of helium's solubility for the full family of alkali metals as well as the introduction of lead (and eventually water) together with lithium as the solvent matrix where helium bubbles will develop, in forward steps towards the modeling of the most realistic environment for the BB of the future fusion reactors.

\acknowledgments{We thank D.Laria, L.A.Sedano and A.Awad for fruitful discussions.  J.M. and L.B. acknowledge financial support from the Generalitat de Catalunya (project "FusionCAT", number J-02603) and from the EUROfusion project (HORIZON-101052200-EUROfusion).  J.M.  and E.A.  thank financial support of project PID2021-124297NB-C32 funded by 
MCIN/AEI/10.13039/501100011033 and “ERDF A way of making Europe” by the “European Union NextGenerationEU/PRTR”.  F.M. acknowledges financial support from the Secretaria d'Universitats i Recerca del Departament d'Empresa i Coneixement de la Generalitat de Catalunya within the ERDF Operational Program of Catalunya (project QuantumCat, Ref.~001-P-001644) and the Spanish MINECO (PID2020-113565GB-C21) funded by MCIN/AEI/10.13039/501100011033.  E.A. thanks to the Spanish Ministry for the Ecological Transition and the Demographic Challenge in relation to the {\it Catedra Argos} grant from the {\it Consejo de Seguridad Nacional (``Amb el suport de la Universitat Politècnica de Catalunya")}.}


\section*{Supplementary Information}
\appendix

We report here a series of mathematical developments and definitions that can help to complete the indications given in the manuscript. They are: (1) the definition of the CoM; (2) the derivation of the formulas to obtain the interatomic forces from potential models and (3) the method to obtain partial pressures and their derivatives.

\section{Center of mass with periodic boundary conditions}

Notice that the CoM of the cluster is determined snapshot by snapshot using PBC, averaging not over the cartesian coordinates $r_i^k$ ($i$ labels the particle and $k = x,y,z$ the direction) but their transformation into sinusoidal functions: $\displaystyle \hat{\zeta} = \frac{1}{N^p_{\rm He}}\sum_{i=1}^{N^p_{\rm He}} {\sin{\left(\frac{2 \pi r_i^k}{L}\right)}}$ and $\displaystyle \hat{\xi} = \frac{1}{N^p_{\rm He}}\sum_{i=1}^{N^p_{\rm He}} {\cos{\left(\frac{2 \pi r_i^k}{L}\right)}}$, and undoing the transformation,

\begin{eqnarray}
        {r}_{\rm CoM}^k =  \frac{L}{2 \pi} ({\rm atan2}({-\hat{\zeta},-\hat{\xi}}) + \pi),
\end{eqnarray}
where ${\bf r}_{\rm CoM} = ({r}_{\rm CoM}^x,{r}_{\rm CoM}^y,{r}_{\rm CoM}^z)$.

\section{Interatomic forces}

The total force acting on the $i$-th atom is expressed as a sum of pairwise forces ${\bf f}_{ij} = f_{ij} \hat{r}_{ij}$ ($i=1,\dots,N$ and $j=1,\dots,i-1,i+1,\dots,N$). The direction of the forces $\hat{r}_{ij} = \frac{{\bf r}_j - {\bf r}_i}{r_{ij}}$ takes into account PBC.    

\subsection{Lithium-lithium forces}

The force applied over the $i$-th lithium atom ($1 < i < N_{\rm Li}$ by convention) is
\begin{eqnarray} \label{force_Li}
    {\bf F}_i^{\rm Li} &=& 
    -\sum_{j \ne i}^{N_{\rm Li}} \left[ \frac{d \varphi_{ij}}{d r_{ij}} + \left( \frac{d \Phi_i}{d \rho_i} + \frac{d \Phi_j}{d \rho_j}\right)\frac{d \psi_{ij}}{d r_{ij}} \right]\frac{({\bf r}_i-{\bf r}_j)}{r_{ij}} 
    - \sum_{j = N_{\rm Li}+1}^{N_{\rm Li}+N_{\rm He}} \frac{dV^{Li-He}_{ij}}{dr_{ij}} \frac{({\bf r}_i-{\bf r}_j)}{r_{ij}} \\ \nonumber
    &=& \sum_{j \ne i}^{N_{\rm Li}} \left[ \frac{d \varphi_{ij}}{d r_{ij}} + \left( \frac{d \Phi_i}{d \rho_i} + \frac{d \Phi_j}{d \rho_j}\right)\frac{d \psi_{ij}}{d r_{ij}} \right]\frac{({\bf r}_j-{\bf r}_i)}{r_{ij}} 
    + \sum_{j = N_{\rm Li}+1}^{N_{\rm Li}+N_{\rm He}} \frac{dV^{Li-He}_{ij}}{dr_{ij}} \frac{({\bf r}_j-{\bf r}_i)}{r_{ij}} 
\end{eqnarray}

The Belashchenko's EAM lithium-lithium pair potential, embedding function and pair contributions to the electron density are rewritten as
\begin{eqnarray}
    \phi(r) = \left\{
    \begin{array}{cc}
         f_0 + f_1(r_0-r) + f_2(e^{f_3(r_0-r)}-1) & \text{ if } r \leq r_0 \\
         \displaystyle k_0 + \frac{k_1}{r} + \frac{k_2}{r^2} + \frac{k_3}{r^3} + \frac{k_4}{r^4} + \frac{k_5}{r^5} & \text{ if } r>r_0  
    \end{array}\right. \;,
\end{eqnarray}




\begin{eqnarray}
    \Phi(r) = \left\{
    \begin{array}{cc}
        a_i + b_i \left( \rho - \rho_{i} \right) + c_i \left( \rho - \rho_{i} \right)^{2} & \text{ if } \rho_{i-1} \leq \rho < \rho_{i} (i=1,2,3,4,5,6) \\
        a_6 + b_6 \left( \rho - \rho_{6} \right) + c_6 \left( \rho - \rho_{6} \right)^{2} & \text{ if } \rho_6 \leq \rho < \rho_{7} \\
        a_7 + b_7 \left( \rho - \rho_{7} \right) + c_7 \left( \rho - \rho_{7} \right)^{3/2} & \text{ if } \rho \geq \rho_7 \\
    \end{array}\right. \;, 
\end{eqnarray}
and
\begin{eqnarray}
    \psi(r) = p_1 e^{-p_2 r} \;.
\end{eqnarray}
Parameters for these functions are compiled in Table~\ref{tab:BEAM}. 
\begin{table}[]
    \centering
    \begin{tabular}{c|cccc}
         \hline \hline
         $i$ & $\rho_i$ & $a_i$ & $b_i$ & $c_i$ \\\hline
         0 & 0 & - & - & - \\
         1 & 0.350 & -0.800385 & -0.449920 & 11.00 \\
         2 & 0.550 & -0.850369 & -0.449920 & -1.000 \\
         3 & 0.700 & -0.878482 & 0.07580 & 1.750 \\
         4 & 0.840 & -0.887963 & -0.210520 & -1.020 \\
         5 & 0.900 & -0.894474 & -0.006520 & 1.700 \\
         6 & 1.000 & -0.89480 & 0 & 0.0326 \\
         7 & 1.100 & -0.894474 & 0.006520 & 0.000 \\ \hline
        \hline
        ~ & $p_i$ \\\hline
        1 & 3.0511 \\
        2 & $\SI{1.2200}{eV \angstrom^{-1}}$\\ \hline
        \hline
        ~ & $r_i$ & $k_i$ & $f_i$ \\\hline
        0 & 2.45 & $\SI{-1.61539351212}{eV}$ & $\SI{0.252868}{eV}$ & - \\
        1 & - & $\SI{32.9193195820}{eV\angstrom}$ & $\SI{0.15252}{eV \angstrom^{-1}}$ & - \\
        2 & - & $\SI{-245.830404172}{eV\angstrom^2}$ & $\SI{0.38}{eV}$ & - \\
        3 & - & $\SI{840.217873656}{eV\angstrom}^3$ & $\SI{1.96}{\angstrom^{-1}}$ & - \\
        4 & - & $\SI{-1369.38125679}{eV\angstrom}^4$ & - & - \\
        5 & - & $\SI{905.623694715}{eV\angstrom}^5$ & - & - \\\hline\hline
    \end{tabular}
    \caption{Parameters used in Belashchenko's EAM pair potential ($r_0$, $k_i$, $f_i$), embedding energy ($\rho_i$, $a_i$, $b_i$, $c_i$) and electron density ($p_1$, $p_2$)\cite{belashchenko2012embedded}.}
    \label{tab:BEAM}
\end{table}
\subsection{Lithium-helium and helium-helium forces}

The force applied over the $i$-th helium atom ($N_{\rm Li} + 1 < i < N_{\rm Li}+N_{\rm He} = N$) is 
\begin{eqnarray} \label{force_He}
            {\bf F}_i^{\rm He} &=& 
            - \sum_{j=1}^{N_{\rm Li}} \frac{dV^{Li-He}_{ij}}{dr_{ij}} \frac{({\bf r}_i-{\bf r}_j)}{r_{ij}} 
            - \sum_{j \neq i}^{N_{\rm Li}+N_{\rm He}} \frac{dV^{He-He}_{ij}}{dr_{ij}} \frac{({\bf r}_i-{\bf r}_j)}{r_{ij}}
             \\ \nonumber
            &=& \sum_{j=1}^{N_{\rm Li}} \frac{dV^{Li-He}_{ij}}{dr_{ij}} \frac{({\bf r}_j-{\bf r}_i)}{r_{ij}} 
            + \sum_{j \neq i}^{N_{\rm Li}+N_{\rm He}} \frac{dV^{He-He}_{ij}}{dr_{ij}} \frac{({\bf r}_j-{\bf r}_i)}{r_{ij}} .
\end{eqnarray}

The Toennies-Tang-Sheng potential used for the modelling of those interactions that involve helium atoms is

\begin{equation}
            V_{\rm TTS} (r) = V_{\rm short} (r) + V_{\rm long} (r)
\end{equation}
where

\begin{equation} \label{eq_TTSshort}
            V_{\rm short} (r) = E_{\rm h} \frac{Z_A Z_B}{r} \left( 1 + a_1 r + a_2 r^2 + a_3 r^3 \right) e^{-\alpha r}
\end{equation}
and

\begin{eqnarray} \label{eq_TTSlong}
            V_{\rm long} (r) = E_{\rm h} D_{\rm e} \left( 1 - e^{-\alpha r}\right) \left\{ \Lambda r^{\frac{7}{\beta}-1} e^{-\beta r} 
            \right. 
            \\ \nonumber
            \left.
            - \sum_{n=3}^{N_{\rm max}} \left[ 1 - e^{-b r} \sum_{k=0}^{2n} \frac{(br)^k}{k!}\right] \frac{C_{2n}}{r^{2n}} \right \}.
\end{eqnarray}
$E_{\rm h}=\SI{27.211386245988(53)}{eV}$ is the Hartree energy and $a_0=\SI{0.529}{\angstrom}$ the Bohr radius.
The series included in the long-distance range (van der Waals) have 5 terms ($N_{\rm max}=8$).
Location and depth of the wells are:
$R_{\rm e}^{\rm Li-He} = \SI{6.070}{\angstrom}$, 
$D_{\rm e}^{\rm Li-He} = \SI{2.003E-04}{eV}$, 
$R_{\rm e}^{\rm He-He} = \SI{2.968}{\angstrom}$, 
$D_{\rm e}^{\rm He-He} = \SI{9.475E-04}{eV}$ a.u.  
The full list of parameters is given in Table~\ref{parameters}.

\begin{table}
\caption{Parameters of TTS potential models.} 
            \begin{tabular}{c|cc}
            \hline \hline
            ~  & Li-He      & He-He     \\ \hline
            $R_{\rm e}$ (\AA) & 6.070 & 2.968 \\
            $D_{\rm e}$ (eV) & $\SI{2.003E-04}{}$ & $\SI{9.475E-04}{}$\\ \hline
            $\displaystyle a_1 \times \left(\frac{R_{\rm e}}{a_0}\right)$ & 11.49538  & 10.34329  \\
            $\displaystyle a_2 \times \left(\frac{R_{\rm e}}{a_0}\right)^2$ & -25.99090 & -23.68667 \\
            $\displaystyle a_3 \times \left(\frac{R_{\rm e}}{a_0}\right)^3$ & 13.49545  & 12.34334  \\
            $\displaystyle \alpha \times \left(\frac{R_{\rm e}}{a_0}\right)$ & 25.06466  & 22.76733  \\
            $\displaystyle \beta \times \left(\frac{R_{\rm e}}{a_0}\right)$ & 14.57866  & 15.14296  \\
            $\displaystyle \frac{7}{\beta a_0}-1$ & 2.19783   & 1.59236   \\
            $\displaystyle b \times \left(\frac{R_{\rm e}}{a_0}\right)$  & 12.38083  & 13.55060  \\
            $\displaystyle \Lambda \times \left( \frac{E_{\rm h}}{D_{\rm e}} \right) \times \left( \frac{R_{\rm e}}{a_0} \right)^{\frac{7}{\beta a_0}-1}$ & $\SI{2.31539e6}{}$ & $\SI{3.5552e6}{}$  \\
            $\displaystyle C_6 \times \left(\frac{E_{\rm h}}{D_{\rm e}} \right) \times \left(\frac{a_0}{R_{\rm e}} \right)^6 $ & 1.43308   & 1.34992   \\
            $\displaystyle C_8 \times \left(\frac{E_{\rm h}}{D_{\rm e}} \right) \times \left(\frac{a_0}{R_{\rm e}} \right)^8$ & 0.47909   & 0.41469   \\
            $\displaystyle C_{10} \times \left(\frac{E_{\rm h}}{D_{\rm e}} \right) \times \left(\frac{a_0}{R_{\rm e}} \right)^{10}$ & 0.22501   & 0.17155   \\
            $\displaystyle C_{12} \times \left(\frac{E_{\rm h}}{D_{\rm e}} \right) \times \left(\frac{a_0}{R_{\rm e}} \right)^{12}$ & 0.14847   & 0.09557   \\
            $\displaystyle C_{14} \times \left(\frac{E_{\rm h}}{D_{\rm e}} \right) \times \left(\frac{a_0}{R_{\rm e}} \right)^{14}$ & 0.13762   & 0.07170   \\
            $\displaystyle C_{16} \times \left(\frac{E_{\rm h}}{D_{\rm e}} \right) \times \left(\frac{a_0}{R_{\rm e}} \right)^{16}$ & 0.17921   & 0.07244   \\ \hline \hline
            \end{tabular}
            \label{parameters}
\end{table}

\section{Estimation of liquid and gas pressures}

The total pressure of a system of volume $V$ is determined by the virial expression

\begin{equation}
          P = \rho k_B T + \frac{1}{3V} \langle \sum_{i=1}^N {\bf r}_{i} \cdot {\bf F}_{i} \rangle = \rho k_B T + \frac{1}{3V} \langle \sum_{i=1}^N {\bf r}_{ij} \cdot {\bf f}_{ij} \rangle,
                \label{eq_totpres} 
\end{equation}
where the sum stands for both Li and He atoms. Partial contributions can be obtained, for homogeneous systems, separating both Li and He terms, so that $P = P_{\rm Li} + P_{\rm He}$.  However, when two separate phases appear the last consideration is no longer true. We define the volume of a bubble as $V_{\rm bubble} = \frac{4}{3} \pi R_{\rm bubble}^3$, where $R_{\rm bubble}$ is its corresponding radius. Then, we approximate the pressure of the bubble as 

\begin{equation}
          P_{\rm bubble} = \frac{1}{\frac{4}{3} \pi R_{\rm bubble}^3} \left( N_{\rm He} k_B T + \frac{1}{3} \langle \sum_{i=N_{\rm Li}+1}^{N_{\rm Li}+N_{\rm He}} {\bf r}_{i} \cdot {\bf F}_{i} \rangle \right), 
\end{equation}
while the pressure of the liquid is considered as 

\begin{equation}
          P_{\rm LM} = \frac{1}{V} \left( N_{\rm Li} k_B T + \frac{1}{3} \langle \sum_{i=1}^{N_{\rm Li}} {\bf r}_{i} \cdot {\bf F}_{i} \rangle \right), 
\end{equation}
where $V$ is the volume of the simulation box.  Notice that

\begin{equation}
          \langle \sum_{i=1}^{N_{\rm Li}} {\bf r}_{i} \cdot {\bf F}_{i} \rangle = \langle \sum_{i=1}^{N_{\rm Li}} \left(\sum_{j=i+1}^{N_{\rm Li}} {\bf r}_{ij} \cdot {\bf f}_{ij} + {\bf r}_{i} \cdot \sum_{j=N_{\rm Li}+1}^{N_{\rm Li}+N_{\rm He}} {\bf f}_{ij}\right) \rangle 
\end{equation}
and

\begin{equation}
          \langle \sum_{i=N_{\rm Li}+1}^{N_{\rm Li}+N_{\rm He}} {\bf r}_{i} \cdot {\bf F}_{i} \rangle = \langle \sum_{i=N_{\rm Li}+1}^{N_{\rm Li}+N_{\rm He}} \left(\sum_{j=i+1}^{N_{\rm He}} {\bf r}_{ij} \cdot {\bf f}_{ij} + {\bf r}_{i}\cdot\sum_{j=1}^{N_{\rm Li}} {\bf f}_{ij}\right) \rangle
\end{equation}
and finally ${\bf r}_{i}\cdot{\bf F}_{ij}+{\bf r}_{j}\cdot{\bf F}_{ji} = ({\bf r}_{i}-{\bf r}_{j}) \cdot {\bf F}_{ij} = {\bf r}_{ij} \cdot {\bf F}_{ij}$.

\section{Pressure's derivatives}
    
The derivative over the radial distance (minus the force between $i$ and $j$) is expressed, in our case, as

\begin{equation}
             \frac{du(r_{ij})}{dr_{ij}} = \frac{d \varphi_{ij}}{d r_{ij}} + \left( \frac{d \Phi_i}{d \rho_i} + \frac{d \Phi_j}{d \rho_j}\right)\frac{d \psi_{ij}}{d r_{ij}}
\end{equation}
for EAM (Li-Li) forces, while it takes the form of

\begin{equation}
             \frac{du(r_{ij})}{dr_{ij}} = \frac{dV(r_{ij})}{dr_{ij}}
\end{equation}
with

\begin{eqnarray}
            \frac{du(r_{ij})}{dr_{ij}} = \left\{ 
            \begin{array}{cc}
            \frac{d \varphi_{ij}}{d r_{ij}} + \left( \frac{d \Phi_i}{d \rho_i} + \frac{d \Phi_j}{d \rho_j}\right)\frac{d \psi_{ij}}{d r_{ij}} & {\rm if } 1<i, j<N_{\rm Li}   \\
                 \frac{dV_{\rm TTS}(r_{ij})}{dr_{ij}} & {\rm if } N_{\rm Li}+1<i<N {\rm and/or } N_{\rm Li}+1<j<N 
            \end{array}
            \right.
\end{eqnarray}
for pairwise (Li-He and He-He) forces.  $r$ stands for the radial distance to the CoM of He atoms.

\section{Binding energies as a function of helium number}

\begin{figure}
        \centering
        \includegraphics[width=.8\columnwidth]{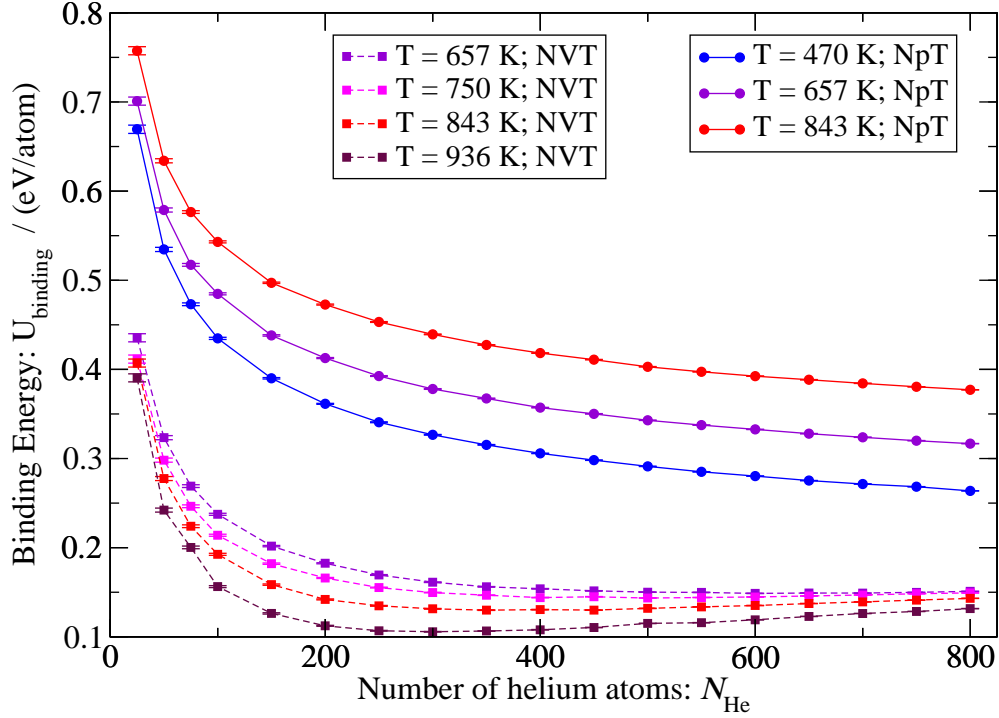}
        \caption{Binding energies as a function of the number of helium atoms}
        \label{fig:Ub_NHe}
\end{figure}

In Fig.\ref{fig:Ub_NHe} the dependence of $U_{\rm binding}$ as a function of the number of helium atoms when simulating the NVT and NpT ensembles is shown.

\section{Further considerations on surface tensions of bubbles}

\begin{table}[]
    \centering
    \begin{tabular}{c|ccc}
         Approach & $A$ (N/m) & $B\times10^4$ (N/m/K) & $R^2$ \\ \hline
         Bakker-Buff & $1.105 \pm 0.003$ & $-5.79 \pm 0.04$ & $0.99990$ \\
         Thompson & $1.263 \pm 0.006$ & $-6.77 \pm 0.08$ & $0.9995$ \\
         Young-Laplace & $1.351 \pm 0.007$ & $-7.35 \pm 0.11 $ & $0.9993$ \\
    \end{tabular}
    \caption{Coefficients of linear regression of the surface tension dependence on the temperature.}
    \label{linfits}
\end{table}

\begin{figure}
            \centering
            \includegraphics[width=0.8\columnwidth]{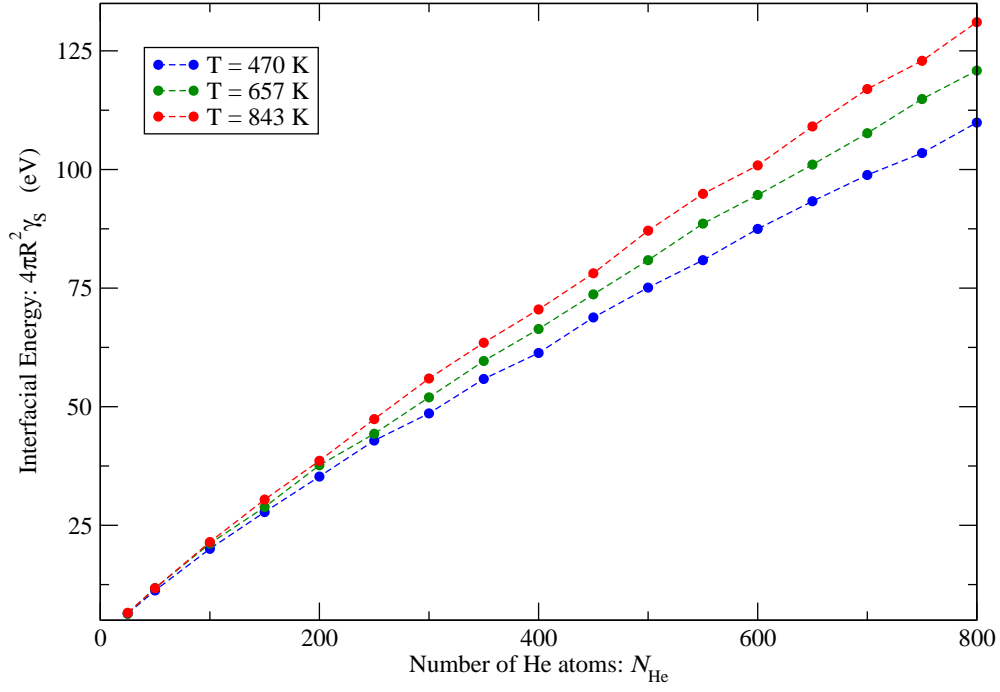}
            \caption{Dependence on their size (number of atoms) of the surface energy of the helium clusters.}
            \label{fig:INTER_ENER}
\end{figure}

The surface energy of a cluster using Eq.~\ref{eq_ST_Bakker} given by 
$\displaystyle 4 \pi \gamma R^2$ is displayed in Fig.\ref{fig:INTER_ENER}.

Notice also that Eqs.~\ref{eq_ST_Thompson} and \ref{eq_ST_Bakker} can be imposed to be equal if a new definition of radius is defined,
\begin{equation} \label{eq_rad}
    R_{\rm S} = \left[ -\frac{\displaystyle -\int_0^\infty r^3 \frac{dp_N(r)}{dr} dr}{\Delta p} \right]^{1/3} . 
\end{equation} 

The two different approaches of the radius calculations are compared in Table~\ref{tab_radius}, being the ones obtained from the mean squared displacement larger than the ones obtained from Eq.~\ref{eq_rad}.
\begin{table}[]
    \centering
    \begin{tabular}{c|cc} \hline \hline
         ~ &  \multicolumn{2}{c}{Radius (\AA)} \\ \cline{2-3}
         $N_{\rm He}$ & $R_{\rm MSD}$ & $R_{\rm S}$ \\ \hline
         100 & 5.596 & 5.028  \\ 
         200 & 7.388 & 6.769 \\
         300 & 8.696 & 8.059 \\
         400 & 9.778 & 9.138 \\
         500 & 10.714 & 10.133 \\
         600 & 11.554 & 10.981 \\
         700 & 12.301 & 11.696 \\
         800 & 12.982 & 12.307 \\ \hline \hline
    \end{tabular}
    \caption{Radii of helium clusters (expressed in \AA) vs $N_{\rm He}$. $R_{\rm MSD}$ corresponds to the radius of gyration, while $R_{\rm S}$ are calculated using Eq.~\ref{eq_rad}.}
    \label{tab_radius}
\end{table}

\newpage
\bibliography{references}


\end{document}